\documentclass[aps,showpacs,preprintnumbers,amsmath,amssymb,pra,twocolumn,superscriptaddress
]{revtex4-1}
\usepackage{txfonts}
\usepackage{float}
\usepackage{graphicx}
\usepackage{dcolumn}
\usepackage{bm}
\usepackage{color}
\usepackage{subfigure}  
\def\comment#1{}

\def\slashchar#1{\setbox0=\hbox{$#1$}           
	\dimen0=\wd0                                 
	\setbox1=\hbox{/} \dimen1=\wd1               
	\ifdim\dimen0>\dimen1                        
	\rlap{\hbox to \dimen0{\hfil/\hfil}}      
	#1                                        
	\else                                        
	\rlap{\hbox to \dimen1{\hfil$#1$\hfil}}   
	/                                         
	\fi}                                         %

\begin{document}

\title{Massless Dirac-like fermions under external fields: a nonminimal coupling approach}

\author{E. Marcelino}
\affiliation{
Instituto de F\'{i}sica, Universidade Federal da Bahia, 40170-115 Salvador, Bahia, Brazil}

\author{E. S. Santos}
\affiliation{
Instituto de F\'{i}sica, Universidade Federal da Bahia, 40170-115 Salvador, Bahia, Brazil}

\author{R. Rivelino}
\affiliation{
Instituto de F\'{i}sica, Universidade Federal da Bahia, 40170-115 Salvador, Bahia, Brazil}

\begin{abstract}
We investigate the unusual properties of quasirelativistic massless fermions under a magnetic or electric field by means of nonminimal couplings. Within this approach, the spin-orbit coupling (SOC) effects are properly generated in the energy spectrum of the quasiparticles. By including a magnetic field, $\textit{B}$, we show that the spin splitting of Landau levels (LLs) obeys a $\sqrt{ B }$ linear dependence with SOC, typical of relativistic particles. Moreover, our calculated spectrum of LLs resembles the behavior of the three-dimensional (3D) massless Kane fermions. Using a nonminimal coupling with an external electric field, we demonstrate that a Rashba-like SOC naturally appears into the relativistic equations and apply to the case of two-dimensional (2D) massless Dirac fermions. Still considering our proposed approach, the Hall conductivity is also computed for the 2D case under transverse electric field both at zero and finite temperatures for a general chemical potential. The results feature a typical quantization of the Hall conductivity at low temperatures, when the absolute value of the gap opened by the electric field is larger than the considered chemical potential. 
\end{abstract}

\maketitle

\section{Introduction}

The discovery of a wide range of new materials for which the electrons obey linear low-energy fermionic excitations \cite{zoo,zoo1}, instead of to obey the Schr\"odinger Hamiltonian, has opened up interesting possibilities to attain faster and more efficient electronics \cite{Castro-Neto,Mota}. These unusual condensed-matter systems share fundamental similarities; e. g., exhibit the so-called Dirac points, or even lines, in their Brillouin zones \cite{Wehling}. For this reason, they are termed Dirac materials \cite{Wehling}, which are completely different from conventional metals and semiconductors \cite{zoo,zoo1}. As important characteristic features of Dirac materials are their response to impurities, magnetic and electric fields, electronic and thermal transport properties, suppression of electron backscattering, optical conductivity, among other unusual properties \cite{DasSarma1}. A significant universal property in the low-energy limit is the power-law temperature dependence of the fermionic specific heat, controlled only by the dimensionality of the excitation phase space \cite{Wehling}.

In the context of Dirac-like materials, Xu et al. \cite{Xu} have recently reported the experimental realization of a Weyl fermion semimetal in single crystalline TaAs, an elusive relativistic massless particle proposed in 1929 \cite{Weyl}. A Weyl semimetal exhibits bulk conduction and valence bands touching at an even number of momentum points with linear dispersion and also belongs to the class of topological materials \cite{Kane_fermions,Bansil}, beyond topological insulators \cite{TI}. Indeed, the presence of a significant spin-orbit coupling (SOC) in certain Dirac materials might induce topological states \cite{Konig}, which exhibit an insulating bulk with a small energy gap and low-energy one-dimensional Dirac edge excitations. Even in the case of pristine graphene, exhibiting only a weak SOC \cite{Graphene}, it has been observed that in the presence of a transverse external electric field, the spin degeneracy is lifted by the Rashba effect \cite{Rashba}, giving rise to interesting band structure topologies \cite{graphenespin}. The effect of an external electric field has also played an important role in the quantum spin/valley Hall conductivity of silicene and its derivatives \cite{Ezawa,Liu,Padilha,XingTao,Tahir}.

A standard approach for studying theories, in which a field under global invariance interacts with a field with local gauge invariance, consists of introducing a proper covariant derivative that couples the current from the global invariant field to the local gauge-invariant field, in such a way that the whole theory becomes local gauge-invariant. This procedure is easy to be performed, which is the so-called minimal coupling and is amenable to a proper geometrical interpretation as suggested by the name "covariant derivative. In this sense, minimal couplings became a standard procedure to introduce interactions in both fundamental and effective theories. Quantum electrodynamics (QED) is the most basic and fundamental theory from the standard model (SM), exhibiting a minimal coupling in which the Dirac's field is coupled to Maxwell's field exhibiting a U(1) local gauge-invariant theory. With the work of Yang and Mills \cite{YM}, the principle of gauge invariance became a paradigm for the description of a complex zoo of particles exhibiting different types of symmetries and conserved charges in the SM, and also in many applications of condensed matter physics.

Despite the success of the minimal coupling procedure, there are other mechanisms to introduce interactions and couple different fields in these theories, the nonminimal couplings \cite{Gazzola}. These could be used, for example, to express the magnetic dipole interaction in Dirac fermions coupled to an electromagnetic field, albeit this idea is generally discarded due to renormalization arguments \cite{Maggiore}. However, for many systems it can be very useful to consider nonminimal couplings, such as in Refs. \cite{Moshinsky,Rivelino,Esdras}, especially for systems subjected to an external classical field \cite{Belissard1,Belissard2}. In particular, nonminimal coupling terms of fermions have been investigated in order to realize a Lorentz-violating background with topological implications \cite{Helayel}. Interestingly, very recently it has been discovered \cite{Hasan} in the context of condensed matter physics a Lorentz-violating Weyl fermion semimetal state in LaAlGe materials.

In this paper, we propose nonminimal couplings to investigate the unusual properties of 2D or 3D relativistic massless fermions under an external electric or magnetic field, taking into account the spin-orbit effects in their energy spectra. We suggest scenarios within this approach, where one can assess the universality of response functions and susceptibilities of Dirac-like systems. Rigorously, to describe the properties of these systems it is required first-principles computational simulations \cite{Rafael} combined with a parametrized effective model Hamiltonian \cite{Kane_fermions,Katsnelson}. However, it is still a great challenge to accurately calculate all the universal properties of Dirac materials. In this sense, our scheme allows a simple way to realize the universality of the unusual properties of the Dirac-like materials without including the effects of many-body interactions in the motion equations. Moreover, some of our results are valid for an arbitrary number of dimensions.

In Sec. II, we obtain the Landau levels (LLs) spectrum for massless fermions coupled to a magnetic field and containing the SOC component. We show that in the 3D case the results resemble the LLs spectrum for Kane fermions in a zinc-blend crystal \cite{Kane_fermions}. In Sec. III, we demonstrate that the SOC term can be directly obtained from a nonminimal coupling, by including an external electrical field. We show that, by reducing to the 2D case, it corresponds to a Rashba-like SOC term in the relativistic description, which may be useful to understand SOC in locally polarized domains resulting from spin recombination channels. In Sec. IV, we consider the nonminimal coupling, as defined in Sec. III for two spatial dimensions, and compute the Hall conductivity at zero and finite temperatures. In Sec. V, we summarize this study considering important features related to the universal properties of Dirac-like materials within different perspectives. 



\section{Landau levels spectrum with SOC effects}

Recently, it has been experimentally demonstrated the existence of condensed-matter systems exhibiting electronic states with conical dispersion in all three dimensions \cite{Kane_fermions,schwarzite}. In particular, electrons in zinc-blend crystal HgCdTe, at the point of the semiconductor-to-semimetal topological transition behave as 3D massless Kane fermions with Fermi velocity about $10^6$ m/s \cite{Kane_fermions}. It is known in this case that in a magnetic field $\textit{B}$ the spin splitting of Landau levels (LLs) follows a $\sqrt{ B }$ linear dependence typical of relativistic particles. 

In this section we obtain a general LLs energy spectrum of a quasirelativistic gas of massless fermions under an external magnetic field, $B$, by using a nonminimal coupling that leaves Eq. (\ref{Free}) linear in both momenta and coordinates, giving rise to a Dirac oscillator \cite{Moshinsky}, which allows to introduce the cyclotron mass effect for the quasiparticles. We have obtained a similar spectrum for the 2D case \cite{Rivelino}, but here we address the case for an arbitrary number of spatial dimensions. 

Firstly, let us consider 3D quasirelativistic massless fermions obeying the Dirac equation in the limit of vanishing mass, which is formally invariant under parity transformation \cite{Semenoff},
\begin{equation} \label{Free}
\gamma^{\mu} \partial_{\mu} \psi =0,
\end{equation}
where the space-time four-vector is defined as $x^{\mu}=(v_Ft,x^1,x^2,x^3)$, the metric tensor is given by $\eta^{\mu \nu}=\eta_{\mu \nu}=diag(1,-1,-1,-1)$, $v_F$ is the Fermi velocity and $\gamma^{\mu}$ are the Dirac matrices, which satisfy the Clifford algebra represented by 
\begin{equation} \label{Clifford}
\{\gamma^{\mu},\gamma^{\nu} \}=2 \eta^{\mu \nu}.
\end{equation}
We also consider sum over repeated indices and that Greek indices belong to the set $\{0,1,2,3\}$ but Latin indices to $\{1,2,3\}$. In the following we use the correspondence $i \partial_{\mu} \rightarrow p_{\mu}=(\frac{\epsilon}{v_F},-\vec{p})$ in Eq. (\ref{Free}) and introduce the nonminimal coupling
\begin{equation}
\vec{p} \rightarrow \vec{p}+ieB\vec{r} \gamma^{0},
\end{equation}
thus, we obtain
\begin{equation}
\gamma^{0} \frac{\epsilon}{v_F} \psi=\vec{\gamma}\cdot (\vec{p}+ieB\vec{r} \gamma^{0})\psi.
\end{equation}
As one can see from the previous equations, we considerd $\hbar=1$, this will still be valid for the rest of this paper. Using Eq. (\ref{Clifford}) and the last equation, together with some manipulations, it is straightforward to obtain
\begin{equation}
\frac{\epsilon^2}{v_{F}^{2}} \psi=\left[ \vec{p}^{2}+e^{2}B^{2}\vec{r}^{2}+deB\gamma^{0}+ieB \gamma^{i} \gamma^{j} (x_{i} p_{j}-x_{j} p_{i})  \gamma^{0} \right] \psi,
\end{equation}
where $d$ is the number of spatial dimensions and the canonical commutation relation, $[x_{i},p_{j}]=i\delta_{ij}$, has been used.
Using the angular momentum $L_{k}=\epsilon_{kij} x^{i}p^{j} \Rightarrow x^{i}p^{j}-x^{j}p^{i}=\epsilon_{kij}L_{k}$ and the definition of the spin operator $S_{k}=\frac{i}{8} \epsilon_{kij} [\gamma^{i} , \gamma^{j}]$, it  is straightforward to obtain
\begin{equation} \label{Magnetic}
\frac{\epsilon^2}{v_{F}^{2}} \psi=\left[ \vec{p}^{2}+e^{2}B^{2}\vec{r}^{2}+eB (d+4 \vec{S} \cdot \vec{L} \gamma^{0} ) \right] \psi,
\end{equation}

Considering $\psi^{T}=(\psi_{1}^{T} \quad \psi_{2}^{T})$ as being a bispinor, we employ the following representation for the $\gamma$ matrices:
\[
\gamma^0=
  \begin{bmatrix}
    I & 0 \\
    0 & -I
  \end{bmatrix}
  \qquad
\gamma^j=
  \begin{bmatrix}
    0 & \sigma^j \\
    -\sigma^j & 0
  \end{bmatrix}
\]
where the $\sigma^{j}$ are the usual Pauli matrices. This is the same representation used in \cite{Moshinsky}, under this representation both $\gamma^{0}$ and the spin operator are diagonal on the Pauli matrices, thus Eq. (\ref{Magnetic}) becomes
\begin{equation} \label{Magnetic_result}
\frac{\epsilon^2}{v_{F}^{2}} \psi_{k}=\left[ \vec{p}^{2}+e^{2}B^{2}\vec{r}^{2} + (-1)^{k+1} 2eB \left(\frac{d}{2} + 2 \vec{S} \cdot \vec{L} \right) \right] \psi_{k}.
\end{equation}
To solve (\ref{Magnetic_result}) one should notice that it can be written in the form, 
\begin{equation} 
\frac{\epsilon^2}{2mv_{F}^{2}} \psi_{k}=\left[ H_{HO} + (-1)^{k+1} \omega \left(\frac{d}{2} + 2 \vec{S} \cdot \vec{L} \right) \right] \psi_{k},
\end{equation}
where we employ the cyclotron frequency $\omega=\frac{eB}{m}$ and $H_{HO}$ is the Hamiltonian operator of the harmonic oscillator;
\begin{equation}
H_{HO}=\frac{\vec{p}^{2}}{2m}+\frac{m\omega^{2}}{2}\vec{r}^{2}.
\end{equation}

Conservation of angular momentum for the harmonic oscillator is truth not only in classical but also in quantum mechanics. In the latter case it is expressed by $[H_{HO},\vec{L}]=0$ (see, e.g., Ref. \cite{Cohen}), so that it is possible to find eigenstates of both $H_{HO}$ and $\vec{S} \cdot \vec{L}$ simultaneously. Since these eigenvalues are already known, it is straightforward to obtain the LLs spectrum of the system;
\begin{equation} \label{Spectrum(B)}
\epsilon^{(k)}_{n}=v_{f} \sqrt{(2n+d) eB +(-1)^{k+1} 2eB \left(\frac{d}{2} + 2 \vec{S} \cdot \vec{L} \right)},
\end{equation}
where $n \in \mathcal{Z}$, corresponds to the LL indices and the index $k \in \{ 1,2 \}$ refers to the components of the bispinor. Thus, we can compute the SOC effects for the two cases of spins up and down, finding the splitting of the energy levels. 

To illustrate this result, we consider the quantum oscillations of 3D massless fermions in a magnetic field. In this case the quasiparticle current loop experiences a torque, which makes the angular momentum $\vec{L}$ precessing around the direction of the magnetic field. However, as demonstrated elsewhere \cite{Rivelino}, the conserved quantity is the total angular momentum  $\vec{J} = \vec{L}+\vec{S}$. Hence, to calculate the LL with SOC effects, we employ $2 \vec{S} \cdot \vec{L}=\vec{J}^{2}-\vec{L}^{2}-\vec{S}^{2}$, which assumes the following values in the 3D case;
\begin{equation}
2 \vec{S} \cdot \vec{L}=
\begin{cases}
l & (j=l+1/2)   \\
-(l+1) & (j=l-1/2).
\end{cases}
\end{equation}

\begin{figure}[H]
	\centering    
	\includegraphics[width=0.51\textwidth]{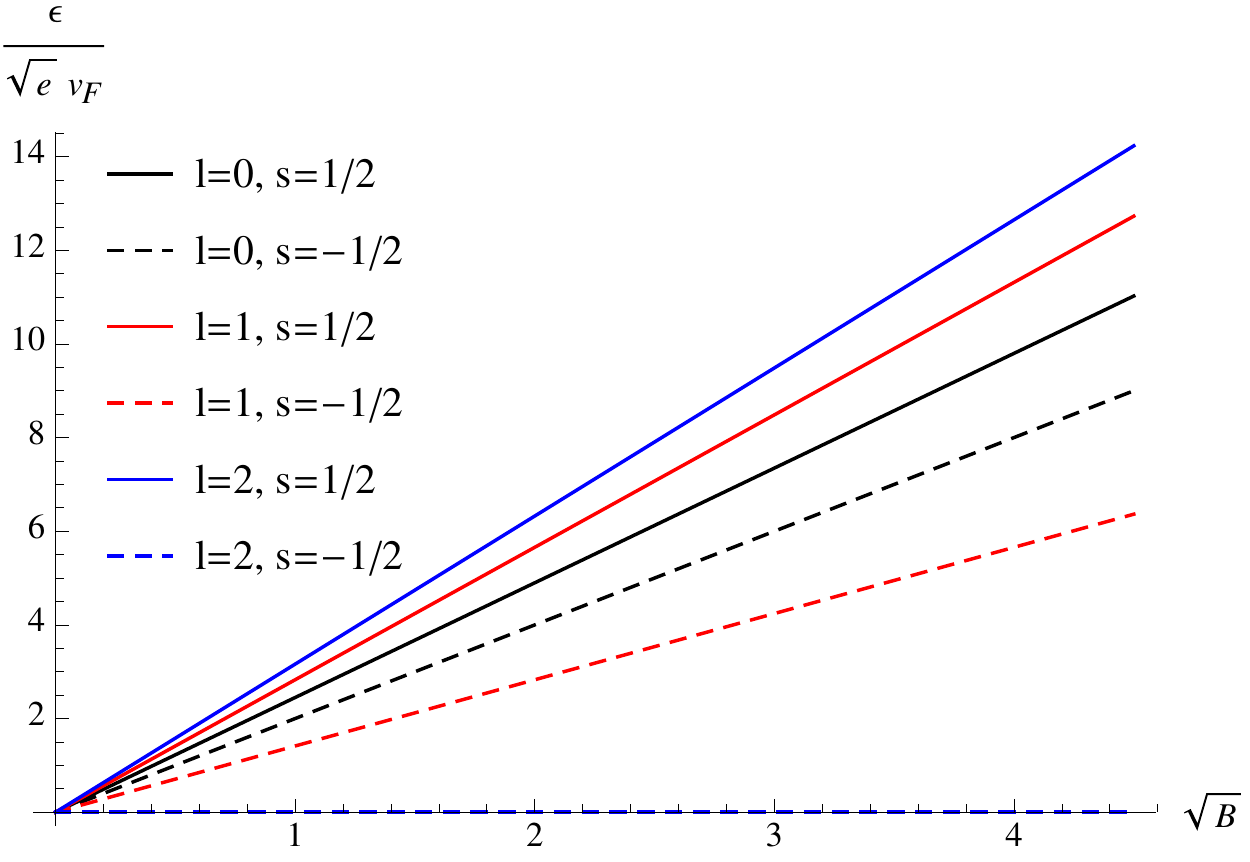}
	\caption{(Color online) Energy spectrum as a function of the square root of the magnetic field for $n = 0$ and different values of $l$. Full lines correspond to $s = 1/2$ and the dashed ones to $s = -1/2$. Dashed blue line corresponding to $l = 2$ and $s = -1/2$ is laying in the horizontal axis, according to Eq. (\ref{Spectrum(B)}).}
	\label{Energy(B)}
\end{figure}

As is known, SOC effects are essential in 3D Dirac semimetals \cite{Gibson}, mainly because of the high atomic weight of the atoms in question, but also in a 3D carbon-based gyroidal schwarzite structure \cite{schwarzite}. For this reason, we compute the energy spectrum with Eq. (\ref{Spectrum(B)}) for different angular momentum quantum numbers and spin projections.
The energy as a function of the magnetic field is displayed in Fig. \ref{Energy(B)} for $k=1$ and in the lowest LL ($n = 0$). We have chosen $k=1$ in Eq. (\ref{Spectrum(B)}) since the $k=2$ case is in general related to other choices of the quantum numbers in the $k=1$ solution. We also claim that the angular momentum quantum numbers $l$ and $s$ might be restricted in order to ensure that the eigenvalues of the Hamiltonian are real numbers.
The present results indicate that for 3D massless Dirac fermions the spin-orbit splitting of LLs follows a $\sqrt{ B }$ dependence typical of relativistic particles. Furthermore, the spin-orbit interaction splits the $2(2l+1)$-fold degenerate valence band in two subspaces corresponding to the total angular momenta $j=l \pm 1/2$. Although we are considering a very simple model of massless 3D fermions, without including interactions, in some extent, this linear dispersion relation is similar to the one observed for massless Kane fermions in a 3D zinc-blend crystal \cite{Kane_fermions}.

\section{Rashba-like coupling in the massless Dirac equation}

Spin-orbit interaction is a fundamental mechanism \cite{DasSarma} that under some broken symmetry can induce spontaneous spin polarization even in the absence of external magnetic fields \cite{Rashba}. This interaction is essential to characterize topological quantum phases of matter as well as to investigate Rashba-like spin splitting effects in 3D Rashba materials \cite{Barone,OMHPs,BiTeI}. Furthermore, considering graphene, which exhibits a weak SOC, in the presence of a transverse external electric field, the spin degeneracy is lifted by the Rashba effect \cite{Rashba}, giving rise to interesting band structure topologies \cite{graphenespin}. The effect of an external electric field has also played an important role in Hall conductivity of silicene \cite{Ezawa}, a monolayer of silicon atoms forming a 2D honeycomb lattice and obeying linear low-energy fermionic excitations with relatively large spin-orbit interaction.  

In this section we introduce a prescription of how to generate a Rashba-like SOC in the quasirelativistic massless Dirac equation, Eq. (\ref{Free}), by using the nonminimal coupling similar to the one proposed in Ref. \cite{Rivelino};



\begin{equation}
\vec{p} \rightarrow \vec{p}+ie \frac{\vec{E}}{\omega} \gamma^{0},
\end{equation}
which leads to the following differential equation
\begin{equation} \label{Linear_E}
\gamma^{0} \frac{\epsilon}{v_F}=\vec{\gamma} \cdot \left( \vec{p}+ie\frac{\vec{E}}{\omega} \gamma^{0} \right) \psi.
\end{equation}
Multiplying this equation by $\gamma^{0}$ on the lhs and squaring the operator acting on $\psi$ on each side, it is simple to obtain
\begin{equation}
\frac{\epsilon^2}{v_{F}^{2}} \psi=\left[ (\gamma^{0} \vec{\gamma}.\vec{p})^{2}-\frac{e^{2} (\vec{\gamma}. \vec{E})^{2}}{\omega^2}- ie\omega \gamma^{0} \gamma^{i} \gamma^{j} (p_{i} E_{j}-E_{i} p_{j}) \right] \psi.
\end{equation}
The first two terms in the rhs of this equation can be simplified by using the Clifford algebra (\ref{Clifford}) and the last one can be written using a commutator of $\gamma$ matrices and Levi-Cevita tensor because of the antisymmetry of the tensor $(p_{i} E_{j}-E_{i} p_{j})$, thus we find
\begin{equation}
\frac{\epsilon^2}{v_{F}^{2}} \psi=\left[ \vec{p}^{2}+\frac{e^{2} \vec{E}^{2}}{\omega^2}- \frac{ie\omega}{4} \epsilon_{kij} [\gamma^{i} , \gamma^{j}] (p_{i} E_{j}-E_{i} p_{j})  \gamma^{0} \right] \psi.
\end{equation}
Using the spin operator defined in Sec. II, and assuming a 3D intrinsic angular momentum representation \cite{Regan}, the last equation can be written as;
\begin{equation} \label{Rashba-eletric}
\left[\frac{\epsilon^2}{v_{F}^{2}} - \vec{p}^{2}-\frac{e^{2} \vec{E}^{2}}{\omega^2}\right] \psi=\frac{4e}{\omega} \vec{S}\cdot(\vec{E} \times \vec{p})  \gamma^{0} \psi.
\end{equation}
We notice that the term in the rhs of Eq. (\ref{Rashba-eletric}) corresponds to a Rashba-like SOC, which can be also obtained with the usual minimal coupling, as demonstrated in Ref. \cite{Rivelino} for a 2D gas of massless Dirac fermions.

Considering the same representation of the $\gamma$ matrices given in Sec. II, the bispinor $\psi^{T}=(\psi_{1}^{T} \quad \psi_{2}^{T})$, and an uniform transverse electric field, $\vec{E}=(0,0,E_{z})$, Eq. (\ref{Rashba-eletric}) can be written in the form;
\begin{equation} \label{Eletric_result}
\frac{\epsilon^2}{v_{F}^{2}} \psi_{k}=\left[ \vec{p}^{2}+ \frac{e^{2} \vec{E}^{2}}{\omega^2} +(-1)^{k+1}\alpha (\sigma_{y} p_{x}-\sigma_{x} p_{y}) \right] \psi_{k}.
\end{equation}
In this approach, the familiar Rashba term \cite{newSOC} emerges when we consider the spin polarization of 2D massless fermions subject to an external electric field, perpendicular to the plane. It is easy to notice that the parameter $\alpha$ appearing in Eq. (\ref{Eletric_result}) is a function of the external field, i. e., $\alpha=2e\omega^{-1} E_{z}$. It is important to mention here that, with a similar procedure, one can obtain the Rashba dependence on a tilted magnetic field \cite{Rashba2}, obtaining the SOC on the LL spectrum. 


\section{Hall conductivity induced by electric field}

Since there is a considerable and continuous interest in quantum spin and quantum valley Hall effects for 2D materials in the presence of an electric field \cite{Ezawa,Liu,Padilha,XingTao,Tahir}, we analyze the effects of an external transverse electric field applied to a gas of 2D massless fermions and compute the Hall conductivity. Starting with the Hamiltonian $H=v_{F} \gamma^{0} \vec{\gamma} \cdot \vec{p}$ for massless free fermions and introducing the nonminimal coupling $\vec{p} \rightarrow \vec{p}+ie\frac{ \vec{E}}{\omega} \gamma^{0}$, we obtain
\begin{equation} \label{H_Hall}
H=v_{F} \gamma_{0} \vec{\gamma} \cdot \left( \vec{p}+ie\frac{\vec{E}}{\omega} \gamma^{0} \right).
\end{equation}

For working in two dimensions, we consider $\gamma^{\mu}=(\sigma_{z},-i \sigma_{x}, -i\sigma_{y})$, which leads to the following Hamiltonian in terms of Pauli matrices
\begin{equation} \label{H-Hall_Pauli}
H=v_{F} \left[-i \sigma_{z} (\vec{\sigma} \cdot \vec{p})-\frac{\vec{\sigma} \cdot \vec{E}}{\omega} \right].
\end{equation}

To calculate the Hall conductivity, let us consider the transverse component of the electric field, $E_z$, and include the z-component of the nonminimal coupling. Indeed, we could try to naively generalize Eq. (\ref{H_Hall}) for three dimensions including the general nonminimal coupling. However, the 3D case leads to four-dimensional $\gamma$ matrices, which would be redundant, since our system is planar, even considering only a perpendicular component of the electric field to the plane. Furthermore, if one simply generalizes Eq. (\ref{H_Hall}) in order to consider a third component of the momentum, the external field will be completely removed from the problem by a simple substitution of variables consisting of a proper shift on the integral of the Hall conductivity. 

Thus, to take into account the contribution of an electric field with transverse component in Eq. (\ref{H-Hall_Pauli}), we generalize the scalar products to three dimensions and consider $p_z=0$. Proceeding in this way the Hamiltonian operator can be written in the form

\begin{equation} \label{Hamiltonian_Hall}
H=\vec{d}\cdot \vec{\sigma},
\end{equation}
where $\vec{\sigma}=(\sigma_x,\sigma_y,\sigma_z)$ is the 3D vector of Pauli matrices and $\vec{d}$ is given by
\begin{equation} \label{d}
\vec{d}=v_{F} \left(-\left[p_y+\frac{E_x}{\omega} \right],p_x-\frac{E_y}{\omega},-\frac{E_z}{\omega} \right).
\end{equation}
Next we introduce the vector $\hat{d} \equiv \vec{d}/|\vec{d}|$ and define a topological charge $(Q)$ in the following expression,
\begin{equation} \label{Top_charge}
Q_{xy}(\vec{p})=\frac{1}{2 \pi} \hat{d}(\vec{p})\cdot [\partial_{p_{x}} \hat{d}(\vec{p}) \times \partial_{p_{y}} \hat{d}(\vec{p})].
\end{equation}
Finally, the Hall conductivity, $\sigma_{xy}$, can be evaluated using the next equation based on Kubo formula \cite{Kubo};
\begin{equation} \label{Hall}
\sigma_{xy}(T,E_z)=\frac{e^{2}}{2h} \int d^{2} \vec{p} Q_{xy}(\vec{p}) \left[n_{F}(-|\vec{d}|)-n_{F}(|\vec{d}|) \right],
\end{equation}
where we define the Fermi distribution $n_{F}(x)=(e^{\frac{1}{T} \left(x-\mu \right)}+1)^{-1}$ and take the Boltzmann constant as being unity $(k_{B}=1)$. Considering the Hamiltonian (\ref{Hamiltonian_Hall}) with Eq. (\ref{d}), the topological charge is given by
\begin{equation} \label{Q}
Q=\frac{eE_z}{2 \pi \omega \left[ (p_{x}-\frac{eE_{y}}{\omega})^{2} + \left( p_{y}+\frac{eE_{x}}{\omega} \right)^{2} + \frac{e^{2}E^{2}_{z}}{\omega^{2}}\right]^{3/2}}.
\end{equation}

Most of the calculations performed in this section are very similar to those done by one of us in Ref. \cite{Marcelino}, although the ideas are quite distinct. For example, in Ref. \cite{Marcelino} the Hall conductivity depends on the magnetization of a system with short range Hubbard interaction and this magnetization has an intrinsic dynamics described by the saddle-point equation. Here, we simply consider an external electric field acting on an ideal gas of massless fermions in two dimensions, motivated, for example, by the the charge carrier behavior in graphene \cite{Novoselov1,Novoselov2,Zhou}, silicene \cite{Ezawa,XingTao,Tahir}, and gemanene \cite{Cheng,Fazzio}. In the following, we report the calculated Hall conductivity at zero and finite temperature.

\subsection{Hall conductivity at zero-temperature}

\begin{figure}[H]
	\centering    
	\includegraphics[width=0.505\textwidth]{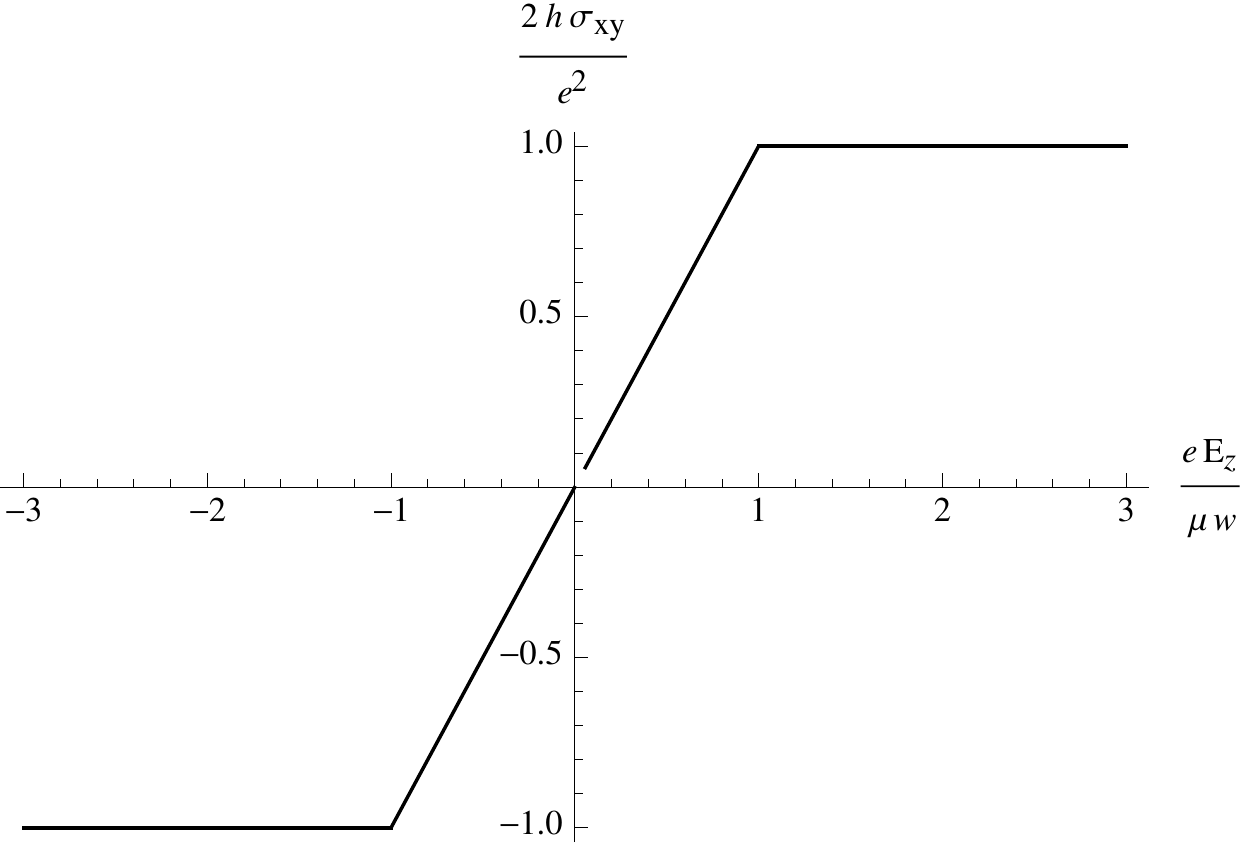}
	\caption{(Color online) Hall conductivity as a function of the ratio between the external electric field and the chemical potential. }
	\label{Hall(T=0)}
\end{figure}

Using Eq. (\ref{Hall}) with the prescription discussed above, we have firstly calculated the Hall conductivity at zero temperature. In this case, the Fermi distributions reduce to Heaviside functions and the Hall conductivity can be explicitly written as
\begin{equation} \label{Sigma(T=0)}
\frac{2 \sigma_{xy}}{h^{-1}e^2}=sign\left(\frac{eE_z}{\omega}\right) \Theta \left( \left|  \frac{eE_z}{\omega \mu} \right|-1\right)+ \frac{eE_z}{\omega \mu} \Theta  \left(1- \left|  \frac{eE_z}{\omega \mu} \right|\right).
\end{equation}
We display the calculated Hall conductivity in Fig. \ref{Hall(T=0)} as a function of the electric field. It is clear from Eq. (\ref{Sigma(T=0)}) and Fig. \ref{Hall(T=0)} that by tuning the transverse electric field, one may tune the Hall effect and unveil topological features of the system.
Our results indicate that for $\left| \frac{eE_{z}}{\mu} \right|$ larger than $\mu$ the Hall conductivity is quantized and may assume two distinct values, the two plateaus in Fig. \ref{Hall(T=0)}. These plateaus correspond to an insulator phase, since the external field is responsible for opening a gap in the ground state energy of the system \cite{graphenespin}. For $\left| \frac{eE_z}{\mu} \right|$ smaller than $\mu$, i.e., in the metallic phase, the Hall conductivity varies linearly between the values from the plateaus in the insulator phase. 

\subsection{Hall conductivity at finite-temperature}

\begin{figure}[H]
	\centering    
	\includegraphics[width=0.5\textwidth]{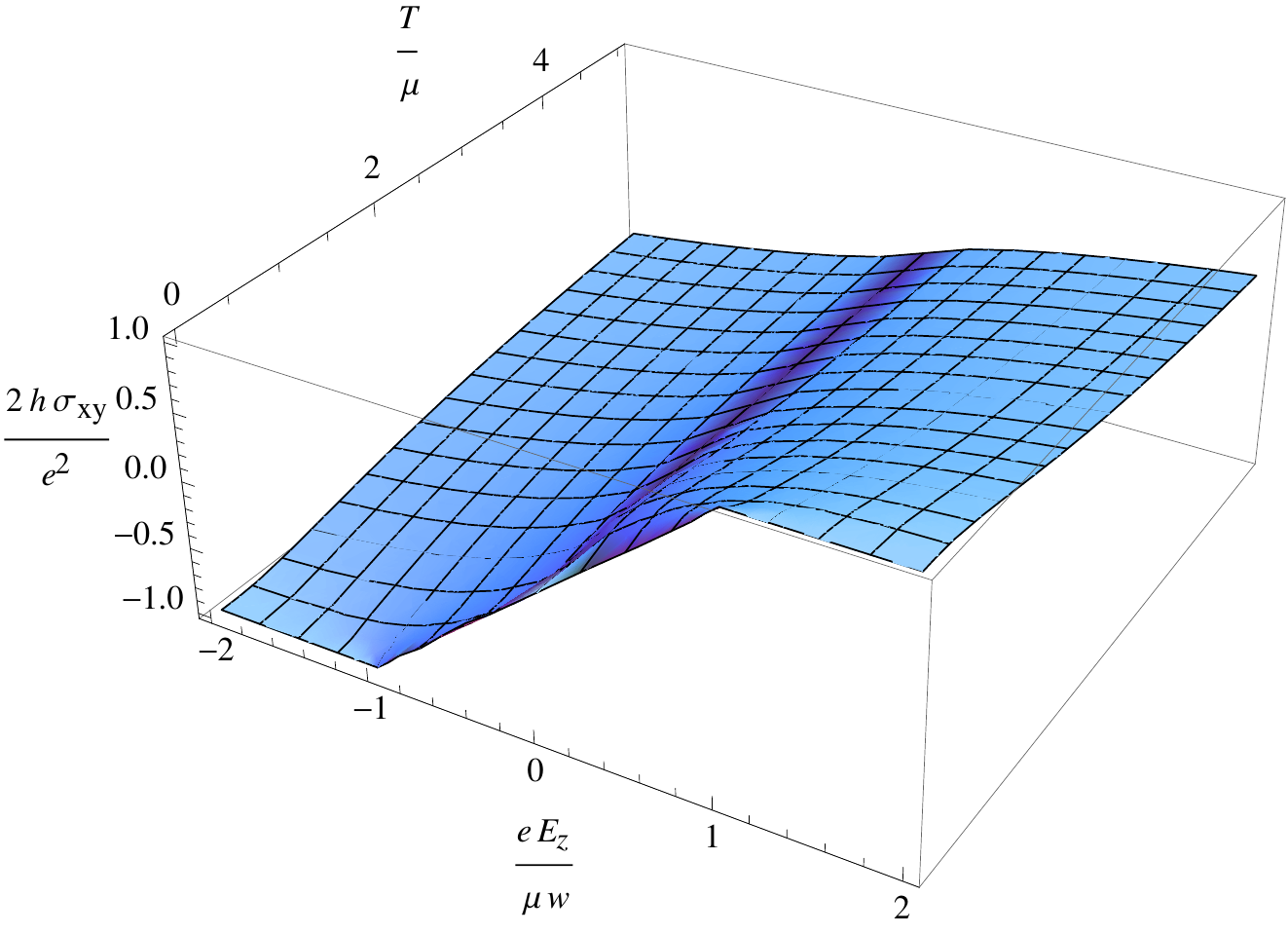}
	\caption{(Color online) Three-dimensional plot of the Hall conductivity as a function of the external field and the temperature. }
	\label{Hall(3D)}
\end{figure}

\begin{figure}[H]
	\centering    
	\includegraphics[width=0.51\textwidth]{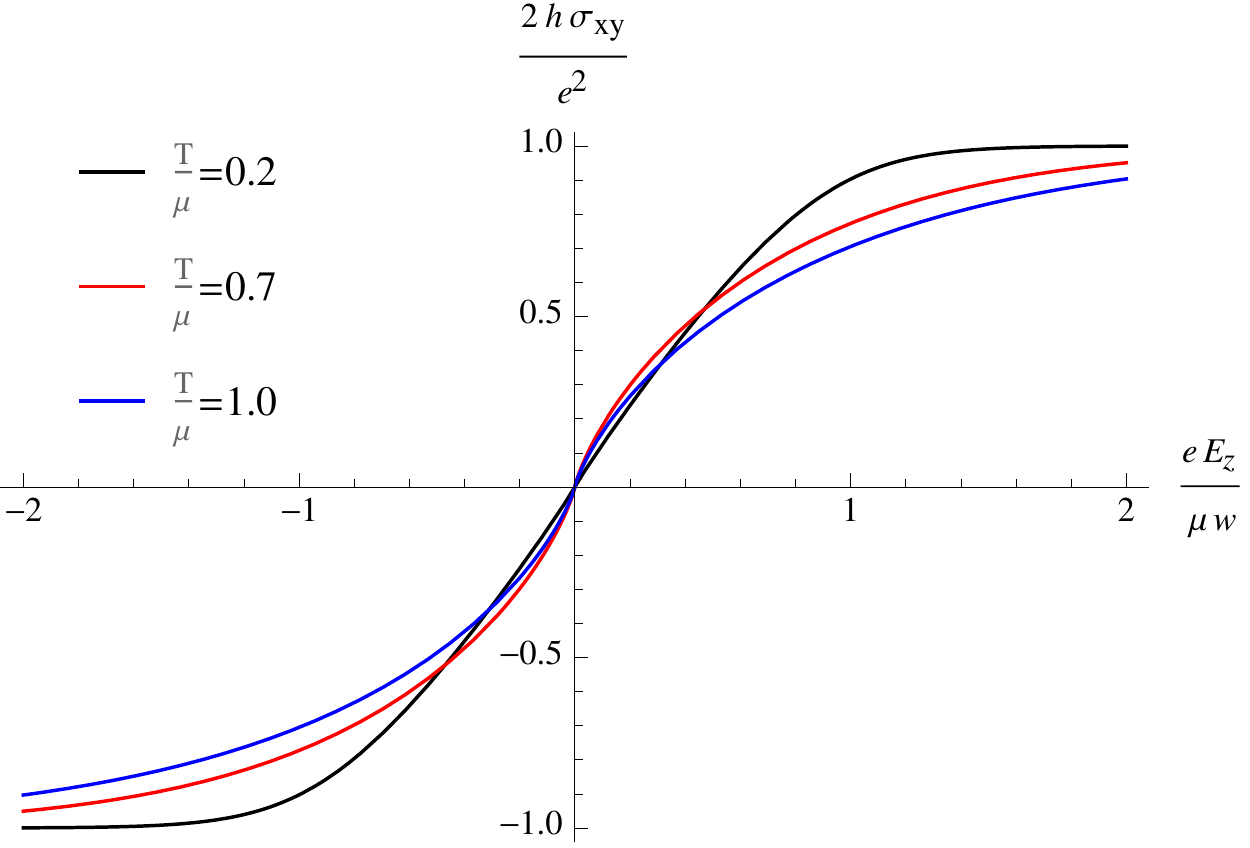}
	\caption{(Color online) Hall conductivity as a function of the field for different values of the temperature. }
	\label{Hall(E)}
\end{figure}

\begin{figure}[h]
	\centering    
	\includegraphics[width=0.5125\textwidth]{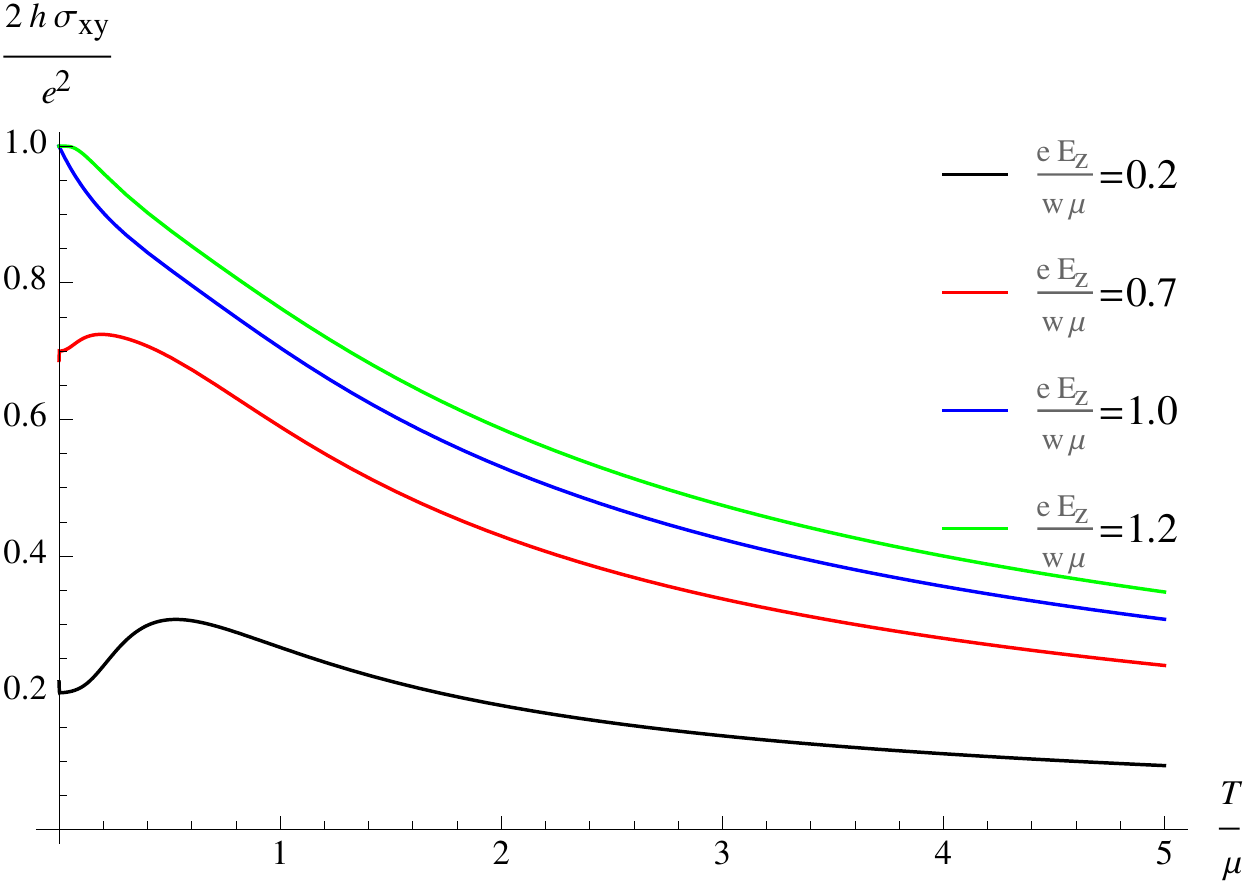}
	\caption{(Color online) Hall conductivity as a function of the temperature for different values of the external field. }
	\label{Hall(T)}
\end{figure}

Since quantum Hall effects can be observed in an experimentally accessible low temperature regime in silicene \cite{Cheng} and germanene \cite{Acun}, we have evaluated the Hall conductivity at finite temperatures. In this case, Eq. (\ref{Hall}) yields
\begin{equation} 
\frac {2h \sigma_{xy}}{e^2} =sign\left(\frac{eE_z}{\omega T}\right)-\frac{eE_{z}}{\omega T} \sum_{\sigma=\pm 1} F_{-1} \left(z^{\sigma},\left| \frac{eE_{z}}{\omega T} \right| \right),
\end{equation}
where $z=e^{\frac{\mu}{T}}$ is the fugacity and we define the incomplete Fermi-Dirac integral in the following form (without multiplying by the Gamma function);
\begin{equation}
F_{\nu}[z,a] \equiv \int_{a}^{\infty} \frac{x^{\nu-1}dx}{\frac{e^{x}}{z}+1}.
\end{equation}
Fig. \ref{Hall(3D)} displays a 3D plot of the Hall conductivity as a function of both the $\frac{eE_z}{ \mu \omega}$ ratio and temperature. As the temperature increases, we observe that the plateaus in the Hall conductivity are smoothed in such a way that the Hall conductivity becomes zero in the $T \rightarrow \infty$ limit, as expected for these types of systems. 
For low temperatures, we notice that the Hall conductivity exhibits a behavior close to that observed in Fig. \ref{Hall(T=0)}, but it changes continuously between the metallic and insulating phases. We display its dependence on the electric field for different temperatures in Fig. \ref{Hall(E)}.
For values of the $\frac{eE_z}{ \mu \omega}$ ratio larger than unity (i.e., in the insulating phase), the Hall conductivity decreases monotonically with the temperature and becomes close to zero in the $T \rightarrow \infty$ limit. When $\frac{eE_z}{ \mu \omega}<1$ (i. e., in the metallic phase), the Hall conductivity increases with the temperature until a certain maximum value and then decreases monotonically, going to zero in the $T \rightarrow \infty$ limit (see Fig. \ref{Hall(T)}). The only vanishing value from our calculated Hall conductivity occurs for $E_z=0$, since the topological charge is zero in this case.

In summary, we have demonstrated in this section that our proposed nonminimal coupling, including a transverse electric field in 2D massless Dirac fermions, leads to a simple and useful Hamiltonian to study quantum Hall effects in these systems in different situations.

\qquad

\section{Conclusions and perspectives}
We have applied a nonminimal coupling approach, via an external electric or magnetic field, to introduce SOC effects in the energy spectra of massless Dirac-like fermions. Our calculations predict, in a first approximation, some unusual aspects of such quantum systems. First, we studied an ideal 3D gas of quasirelativistic massless fermions subject to a magnetic field. Our results have shown the external field induces quantum oscillations and energy spectrum features a linear dependence with the square root of the modulus of the magnetic field, typical of Kane fermions in a zinc-blend crystal. We also considered a nonminimal coupling to introduce an external electric field in the relativistic massless system. In this case, we obtained that (i) the Rashba SOC can be generated for a specific 2D case; (ii) the Rashba parameter was obtained as a function of the external electric field. 

In this approach, we have considered a 2D ideal gas of massless fermions under an an external electric field to study the conduction properties. Based on this model, we calculated the Hall conductivity of the system at zero and finite temperature. Our results indicated that only the transverse component of the external field opens a gap. Moreover, the Hall conductivity is quantized in the insulator region and features a linear dependence on the modulus of the electric field in the metallic region at zero temperature. Interestingly, finite values of the Hall conductivity may be achieved even at large temperatures for large enough fields, although they are always smaller than an absolute value. Notwithstanding, although the Hall conductivity decreases monotonically with the field in the insulator region, it increases with the temperature until a maximum value before it decreases in he metallic region.

Although in the present paper we only considered ideal models, our results suggest scenarios where one can assess the universality of response functions and susceptibilities of Dirac-like systems. Thus, future perspectives include applications to systems featuring interaction and/or disorder, more realistic and sophisticated tight-binding models in the context of Dirac-like materials may also be considered.

\section*{Acknowledgement}
The authors acknowledge financial support from the Brazilian agencies CAPES, CNPq and FAPESB.

\end{document}